\documentclass[10pt,a4paper]{article}

\newcommand{\xa}{\mathbf{x}^a}
\newcommand{\xb}{\mathbf{x}^b}
\newcommand{\xt}{\mathbf{x}^t}
\newcommand{\xx}{\mathbf{x}}
\newcommand{\yo}{\mathbf{y}^o}

\newcommand{\xaref}{\xa_{ref}}

\newcommand{\yoref}{\yo_{ref}}

\newcommand{\oA}{\mathbf{A}}
\newcommand{\oB}{\mathbf{B}}
\newcommand{\oC}{\mathbf{C}}

\newcommand{\oH}{\mathbf{H}}
\newcommand{\oI}{\mathbf{I}}
\newcommand{\oK}{\mathbf{K}}

\newcommand{\oM}{\mathbf{M}}

\newcommand{\oP}{\mathbf{P}}
\newcommand{\oR}{\mathbf{R}}
\newcommand{\oS}{\mathbf{S}}
\newcommand{\oQ}{\mathbf{Q}}

\newcommand{\tP}{\mathbf{\tilde{P}}}
\newcommand{\tR}{\mathbf{\tilde{R}}}
\newcommand{\tS}{\mathbf{\tilde{S}}}
\newcommand{\tQ}{\mathbf{\tilde{Q}}}

\usepackage{graphicx}
\usepackage{dcolumn}
\usepackage{bm}
\begin{document}

\title{Robustness of nuclear core activity reconstruction by data assimilation}

\author{Bertrand Bouriquet $^1$ \footnote{bertrand.bouriquet@cerfacs.fr}
   \and Jean-Philippe Argaud $^{2,1}$
   \and Patrick Erhard $^2$
   \and S\'ebastien Massart $^1$
   \and Ang\'elique Pon{\c c}ot $^2$
   \and Sophie Ricci $^1$
   \and Olivier Thual $^{1,3}$}

\maketitle

\footnotetext[1]{
Sciences de l'Univers au CERFACS, URA CERFACS/CNRS No~1875,
42 avenue Gaspard Coriolis,
F-31057 Toulouse Cedex 01 - France
}
\footnotetext[2]{
Electricit\'e de France,
1 avenue du G\'en\'eral de Gaulle,
F-92141 Clamart Cedex - France
}
\footnotetext[3]{
Universit\'e de Toulouse, INPT, UPS, IMFT,
All\'ee Camille Soula,
F-31400 Toulouse - France
}

\begin{abstract}

We apply a data assimilation techniques, inspired from meteorological
applications, to perform an optimal reconstruction of the neutronic activity
field in a nuclear core. Both measurements, and information coming from a
numerical model, are used. We first study the robustness of the method when the
amount of measured information decreases. We then study the influence of the
nature of the instruments and their spatial repartition on the efficiency of the
field reconstruction.

 
\end{abstract}

\section{Introduction\label{sec:int}} 

In this paper, we focus on the efficiency of a neutronic field reconstruction
procedure with Data Assimilation when varying the number and the repartition of
the available instruments. The data assimilation technique used for this
reconstruction allows to combine, in an optimal and consistent way, information
coming either from measurements or from a numerical model.

Data assimilation methods are not commonly used in nuclear core physics
\cite{Massart07}, contrary to meteorology or oceanography
\cite{Parrish92,Todling94,Ide97}. The procedure proposed here is the same as the
one meteorologists use to obtain high accuracy meteorological reconstructed
fields in time and space. This is the case, for example, of the commonly used
meteorological re-analysis data set ERA-40 \cite{era40} among others
\cite{Kalnay96,Huffman97}. 

One of the main advantages of data assimilation is that it takes into account
every kind of heterogeneous information within the same framework. Moreover,
this method has a formalism that allows to adapt itself to instrument
configuration change.  We exploit this last property here to study the quality
of the reconstructed activity field as a function of the number of available
measurements. A major point in this study is to estimate the instrumented system
robustness in the framework of a data assimilation reconstruction procedure.
Moreover, such a study also informs about the effect of instrumentation design
within a nuclear core and the resilience to instrument removal. 

In this paper, we first detail the data assimilation method and how it addresses
field reconstruction. To evaluate the influence of the number of instruments on
the activity field reconstruction, the repeated application of the method faces
some huge computational issues. Those difficulties are overcomed using a matrix
inversion method based on the Schur complement. A detailed presentation of this
method is presented in Appendix~\ref{sec:schur}. First we present the results on a
standard case with synthetic measurements and comments on them. To get a better
understanding, we extend the results to other instrumental repartitions and
other errors settings. This allows us to give some conclusions on the error and
instrument repartition effects in activity field reconstruction using data
assimilation.  

\section{Data assimilation\label{sec:da}}

We briefly introduce the useful data assimilation key points to understand their
use as applied in \cite{Talagrand97,Kalnay03,Bouttier99}. Data assimilation is a
wider domain and these techniques are, for example, the keys of the nowadays
meteorological operational forecasts \cite{Rabier2000}. This is through advanced
data assimilation methods that weather forecasting has been drastically improved
during the last 30 years. All the available data, such as satellite measurements
as well as sophisticated numerical models, are used. 

The ultimate goal of data assimilation methods is to estimate the inaccessible
true value of the system state,  $\xt$ where the $t$ index stands for "true
state" in the so called "control space". The basic idea is to combine
information  from an \textit{a priori} on the state of the system (usually
called $\xb$, with $b$ for "background"), and  measurements (referenced as
$\yo$). The background is usually the result of numerical simulations, but can
also be derived from any {\it a priori} knowledge. The result of data
assimilation is called the analysis, denoted by $\xa$, and it is an estimation
of the true state $\xt$ we want to approximate.   

The control and observation spaces are not necessary the same, and a bridge
between them needs to be built. This is the observation operator $H$, that
transforms values from the space of the background to the space of observations.
For our data assimilation purpose we will use its linearisation $\oH$ around the
background. The inverse operation going from observation increments to
background increments is given by the transpose $\oH^T$ of $\oH$.

Two other ingredients are necessary. The first one is the covariance matrix of
observation errors, defined as $\oR=E[(\yo-H(\xt)).(\yo-H(\xt))^T]$ where $E[.]$
is the mathematical expectation. It can be obtained from the known errors on
unbiased measurements which means $E[\yo-H(\xt)]=0$. The second one is the
covariance matrix of background errors, defined as
$\oB=E[(\xb-\xt).(\xb-\xt)^T]$. It represents the error on the {\it a priori}
state, assuming it to be unbiaised following the $E[\xb-\xt]=0$ no biais
property. There are many ways to get this {\it a priori} state and background
error matrices. However, those matrix are commonly the output of a model and an
evaluation of accuracy, or the result of expert knowledge. 

It can be proved, within this formalism, that the Best Unbiased Linear Estimator
(BLUE) $\xa$, under the linear and static assumptions, is given by the following
equation:
\begin{equation}\label{xa}
\xa = \xb + \oK \big(\yo- H\xb\big),
\end{equation}
where $\oK$ is the gain matrix:
\begin{equation} \label{Kmat}
\oK = \oB\oH^T (\oH\oB\oH^T + \oR)^{-1}.
\end{equation}
Moreover, we can get the analysis error covariance matrix $\oA$, characterising
the analysis errors $\xa-\xt$. This matrix can be expressed from $\oK$ as:
\begin{equation}
\oA = (\oI - \oK\oH)\oB,
\end{equation}
where $\oI$ is the identity matrix.

It is worth noting that solving Equation \ref{xa} is, if the probability
distribution is Gaussian, equivalent to minimise the following function
$J(\xx)$, $\xa$ being the optimal solution:
\begin{equation}\label{J}
J(\xx) = (\xx-\xb)^T \oB^{-1}(\xx-\xb) \medskip + \big(\yo-\oH\xx\big)^T \oR^{-1} \big(\yo-\oH\xx\big).
\end{equation}

This minimisation is known in data assimilation as 3D-Var methodology
\cite{Talagrand97}.


\section{Data assimilation implementation}

The framework of this study is the standard configuration of a 900 MWe nuclear
Pressurized Water Reactor (PWR900). To perform data assimilation, both
simulation code and data are needed. For the simulation code, the EDF
experimental calculation code for nuclear core COCAGNE in a standard
configuration is used. The description of the basic features of this model are
done in Section \ref{model}.

To have a good understanding of the instrumentation effect, we want to study
various kind of configurations, even some that do not exist operationally and so
cannot be tested experimentally. For that purpose, synthetic data  are used that
allows to have an homogeneous approach all along the document. Synthetic data is
generated from a model simulation, filtered through an instrument model and
noised according to a predefined measurement error density function (usually of
Gaussian type). 

\begin{figure}[!ht]
\begin{center}
  \includegraphics[width=1.0\textwidth]{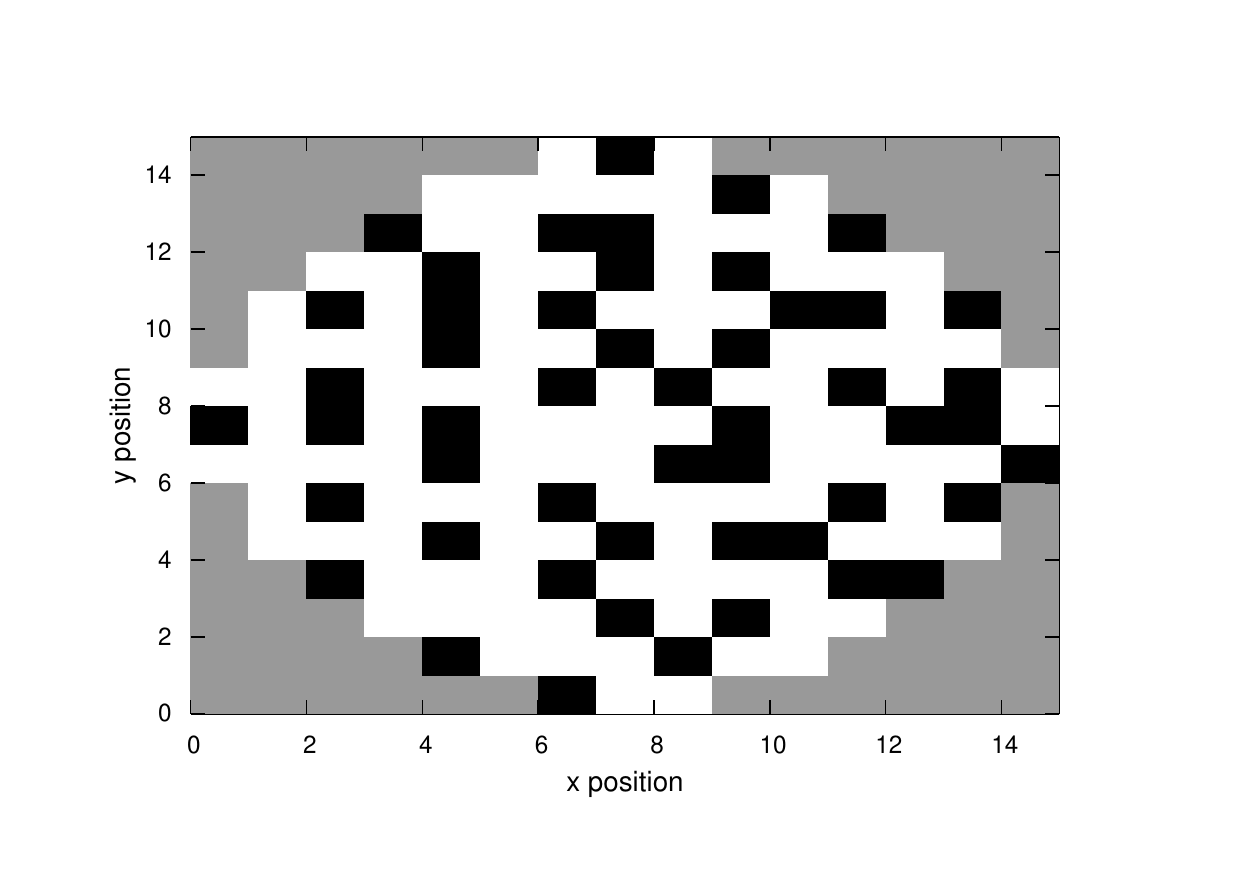}

  \caption{ The positions of MFC instruments in the nuclear core are localised
  in assemblies in black within the horizontal slice of the core. The assemblies
  without instrument are marked in white and the reflector, out of the reactive
  core, is in gray. \label{fig:figMFC}}

\end{center}
\end{figure} 

In the present case, we study the activity field reconstruction. An horizontal
slice of a PWR900 core is represented on the Figure \ref{fig:figMFC}. There is a
total of $157$ assemblies within this core. Among those assemblies, $50$ are
instrumented with Mobiles Fissions Chambers (MFC). Those assemblies are divided
verticaly in $29$ vertical levels. Thus, the size of the control $\xx$ is $4553$
$(157 \times 29)$. The size of the observation vector $\yo$ is $1450$ $(50
\times 29)$. 

\subsection{Brief description of the nuclear core model\label{model}}

The aim of a neutronic code like COCAGNE is to evaluate the neutronic activity
field and all associated values within the nuclear core. This field depend on
the position in the core and on the neutron energy. To do such an evaluation,
the population of neutrons are divided in several groups of energy. In the
present case only two groups are taken into account giving the neutronic flux
$\Phi=(\Phi_1,\Phi_2)$ (even if the present code have no limit for the group
number). The material properties depend on the position in the core, as the
neutronic flux $\Phi$, identified by solving two-group diffusion equations
described by:
\begin{equation}\label{eq:cirep:eqn}
\left\{
\begin{array}{l}
\displaystyle - \mbox{div}(D_1\mbox{\textbf{grad}}\Phi_1) + (\Sigma_{a1} + \Sigma_r ) \Phi_1 = 
\frac{1}{k} \Big(\nu_1 \Sigma_{f1} \Phi_1 + \nu_2 \Sigma_{f2} \Phi_2 \Big) \\
\displaystyle -\mbox{div}(D_2\mbox{\textbf{grad}}\Phi_2) + \Sigma_{a2} \Phi_2 - \Sigma_r \Phi_1 = 0, \\
\end{array}
\right.
\end{equation}
where $k$ is the effective neutron multiplication factor, all the quantities and
the derivatives (except $k$) depend on the position in the core, $1$ and $2$ are
the group indexes, $\Sigma_r$ is the scattering cross section from group $1$ to
group $2$, and for each group, $\Phi$ is the neutron flux, $\Sigma_a$ is the
absorption cross section, $D$ is the diffusion coefficient, $\nu\Sigma_f$ is the
corrected fission cross section.

The cross sections also depend implicitly on the concentration of boron, which
is a substance added in the water used for the primary circuit to control the
neutronic fission reaction, throught a feedback supplementary model. This model
takes into account the temperature of the materials and of the neutron
moderator, given by external thermal and thermo-hydraulic models. A detailed
description of the core physic and numerical solving can be found in reference
\cite{Duderstatd76}.

The overall numerical resolution consists in searching for boron concentration
such that the eigenvalue $k$ is equal to $1$, which means that the nuclear power
production is stable and self-sustaining. It is named critical boron
concentration computation.

The activity in the core is obtained through a combination of the fluxes
$\Phi=(\Phi_1,\Phi_2)$, given on the chosen mesh of the core. Using homogeneous
materials for each assembly (for example $157$ in a PWR900 reactor), and
choosing a vertical mesh compatible with the core (usually $29$ vertical
levels), this result in a field of activity of size $157 \times 29=4553$ that
cover all the core.

\subsection{The observation operator $H$}

The $H$ observation operator is the composition of a selection and of a
normalisation procedure. The selection procedure extracts the values
corresponding to effective measurement among the values of the model space. The
normalisation procedure is a scaling of the value with respect to the geometry
and power of the core. The overall operation is non linear. However, with a
range of value compatible with assimilation procedure, we can calculate the
linear associated operator $\oH$. This observation matrix is a  $(4553 \times
1450)$ matrix.

\subsection{The background error covariance matrix $\oB$}

The $\oB$ matrix represents the covariance between the spatial errors for the
background. In order to get those, we estimate them as the product of a
correlation matrix $\oC$ by a normalisation factor.  

The correlation $\oC$ matrix is built using a positive function that defines the
correlations between instruments with respect to a pseudo-distance in model
space. Positive functions have the property (via Bochner theorem) to build
symmetric defined positive matrix when they are used as matrix generator
\cite{Matheron70,Marcotte08}. In the present case, Second Order Auto-Regressive
(SOAR) function is used to prescribe the $\oC$ matrix. In such a function, the
amount of correlation depends from the euclidean distance between spatial
points. The radial and vertical correlation length ($L_r$ and $L_z$
respectively, associated to the radial $r$ coordinate and the vertical $z$
coordinate) have different values, which means we are dealing with a global
pseudo euclidean distance. The used function can be expressed as follow:
\begin{equation}  
C(r,z) = \left(1+\frac{r}{L_r}\right) \left(1+\frac{|z|}{L_z}\right)
         \exp{\left(-\frac{r}{L_r}-\frac{|z|}{L_z}\right)}.
\label{eqB}
\end{equation}
The matrix obtained by the above Equation \ref{eqB} is a correlation matrix. It
is then multiplied by a suitable variance coefficient to get covariance matrix.
This coefficient is obtained by statistical study of difference between model
and measurements in real case. In our case, the size of the  $\oB$  matrix is
related to the size of model space so it is $(4553 \times 4553)$.

\subsection{The observation error covariance matrix $\oR$}

The observation error covariance matrix $\oR$ is approximated by a diagonal
matrix. This means it is assumed that no significant correlation exists between
the measurement errors of the MFC. The usual modelling is to take those value as
a percentage of the observation. This can be expressed as:
\begin{equation}  
\oR_{jj} = \left( \alpha (y^o)_j \right) ^{2}, \quad \forall j
\label{eqR}
\end{equation}
The parameter $\alpha$ is fixed according to the accuracy of the measurement and
the representative error associated to the instrument. The size of the  $\oR$
matrix is related to the size of observation space, so it is $(1450 \times
1450)$.

\section{General results on instrument removal}

To test the robustness, many BLUE calculations need to be done to evaluate the
results quality with instruments configuration modifications. We want to have an
evaluation of the quality of reconstruction as a function of the number of
instruments, with a significant statistical result. To efficiently perform these
numerous computations, a specific method using Schur complement was developped.
The details of this new method are reported in Appendix~\ref{sec:schur}.

Here, we are interested in the evaluation of the quality of the analysis $\xa$ as
a function of the amount of provided information. To quantify this effect we
make a statistic of $200$ scenarios of instruments removal. We are making those
statistics on several hypothesis, starting from a complete instrument
configuration and then removing instruments two by two until none remains. The
calculation are done on the basis of the algorithm and hypothesis on data
assimilation described previously.

To quantify the impact of removed instruments on the analysis, we look at the
percentage quantity $v$ defined as follow:
\begin{equation} \label{yHxayHxb}
v = 100 \frac{||\yoref-H\xb||-||\yoref-H\xa||}{||\yoref-H\xb||-||\yoref-H\xaref||},
\end{equation}
where $\xaref$ corresponds to the analysis when no instrument is removed (this
is the best estimation possible with respect to the information available on the
system), and where $\yoref$ and $H$ are the reference observations and
observation operator used to build $\xaref$. $H$ stand for the observation
operator when no instrument are removed. This criterium, which is basd on the
norm of the innovation vector $\yoref-H\xb$, focuses on measurements. Since
$||\yoref-H\xa||$ is greater than $||\yoref-H\xaref||$ (best estimate) and
smaller than $||\yoref-H\xb||$ (innovation), $v$ is a measure of the quality of
the analysis.

Such a definition have several advantages. First of all, the limit of this
function are interesting. On the one hand, limit when no instrument is removed
is $100\%$. On the other hand, the limit when all instruments are removed is
$0\%$. With such a formula we can compare the variation of the information on a
unique scale. If we obtain some value above the limit of $100\%$, this mean the
parametrisation of data assimilation was not done correctly.

The interest of using this formula is that it can be applied directly as well to
experimental data as to synthetic data without any change.

\begin{figure}[!ht]
\begin{center}
  \includegraphics[width=1.0\textwidth]{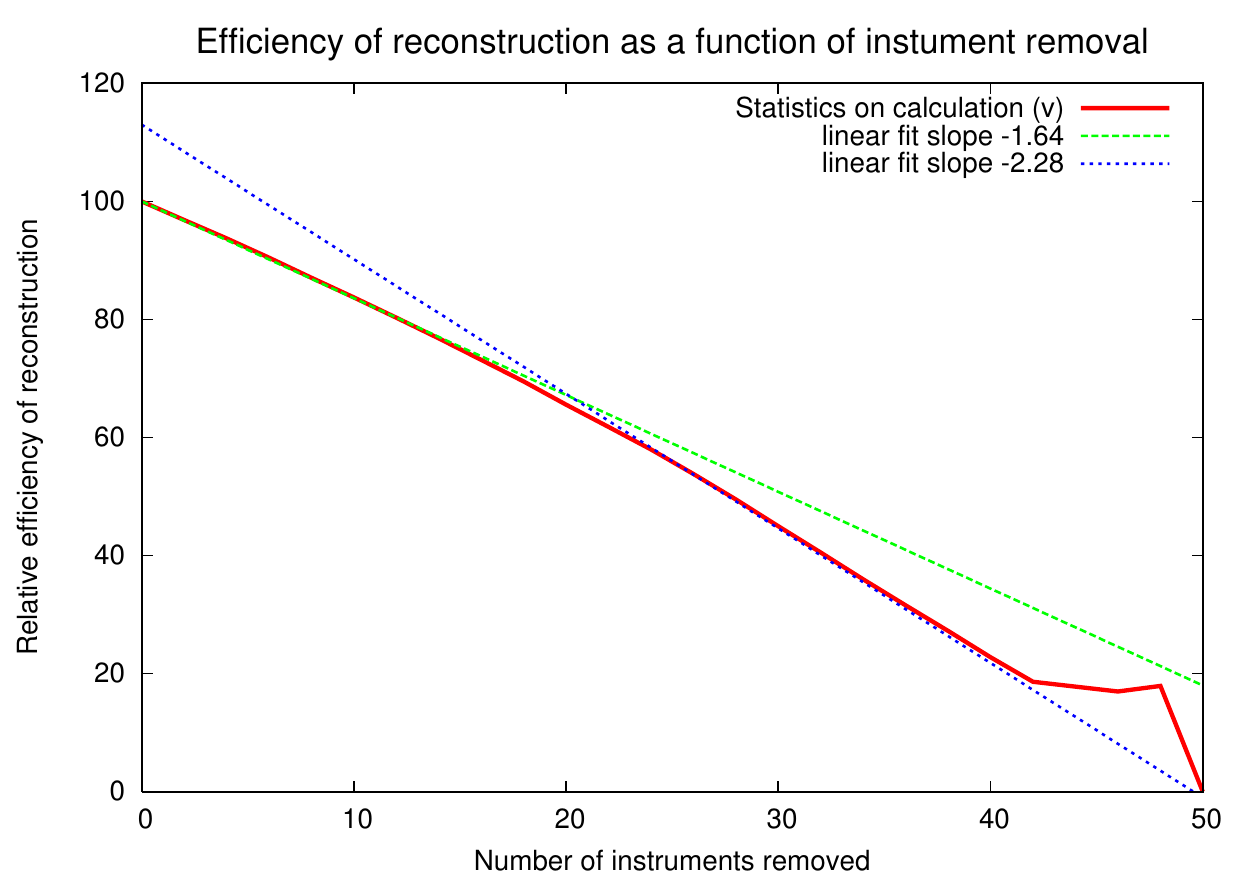}

  \caption{The plain curve represents the value $v$ given by Equation 
  \ref{yHxayHxb} as a function of the number of instruments removed. This
  results come from a typical PWR900 instrument location. The two other lines
  are linear regression corresponding to both decreasing regime.
  \label{fig:pgi-fig1}}

\end{center}
\end{figure}

On Figure \ref{fig:pgi-fig1} are presented the results of the quantity $v$ as
a function of the number of removed instruments.

As expected, the relative quality of reconstruction decreases as a function of
the number of instruments removed. However within this decrease, three phases
can be seen:

\begin{enumerate}

\item  A first phase of slow decreasing until  we  removed roughly $20$
instruments. This phase is rather clear and can be fitted by a linear regression
with a slope of $-1.64$  ({\it arbitrary unit (a.u.) per instrument}). The fit
is shown in (green) dash line on Figure \ref{fig:pgi-fig1}.

\item After $20$ instrument removed the decreasing speed of the slope
increases.  The second linear fit has a slope of $-2.28$ ({\it a.u. per
instrument}). This fit is shown in (blue) dotted line in  Figure
\ref{fig:pgi-fig1}.

\item Beyond $40$ instruments removed, we  reach a third phase of stagnation
then a brutal decrease to $0$ the limit value imposed by Equation
\ref{yHxayHxb}.

\end{enumerate}

This characteristic behaviour can be seen in several cases that we studied. We
have also noticed it on real measurements \cite{SFP07,ecmi2008}. The transition
between the two first decreasing phases is specially strong when we do the
analysis using real measurements.   

\begin{figure}[!ht] 
\begin{center} 
  \includegraphics[width=1.0\textwidth]{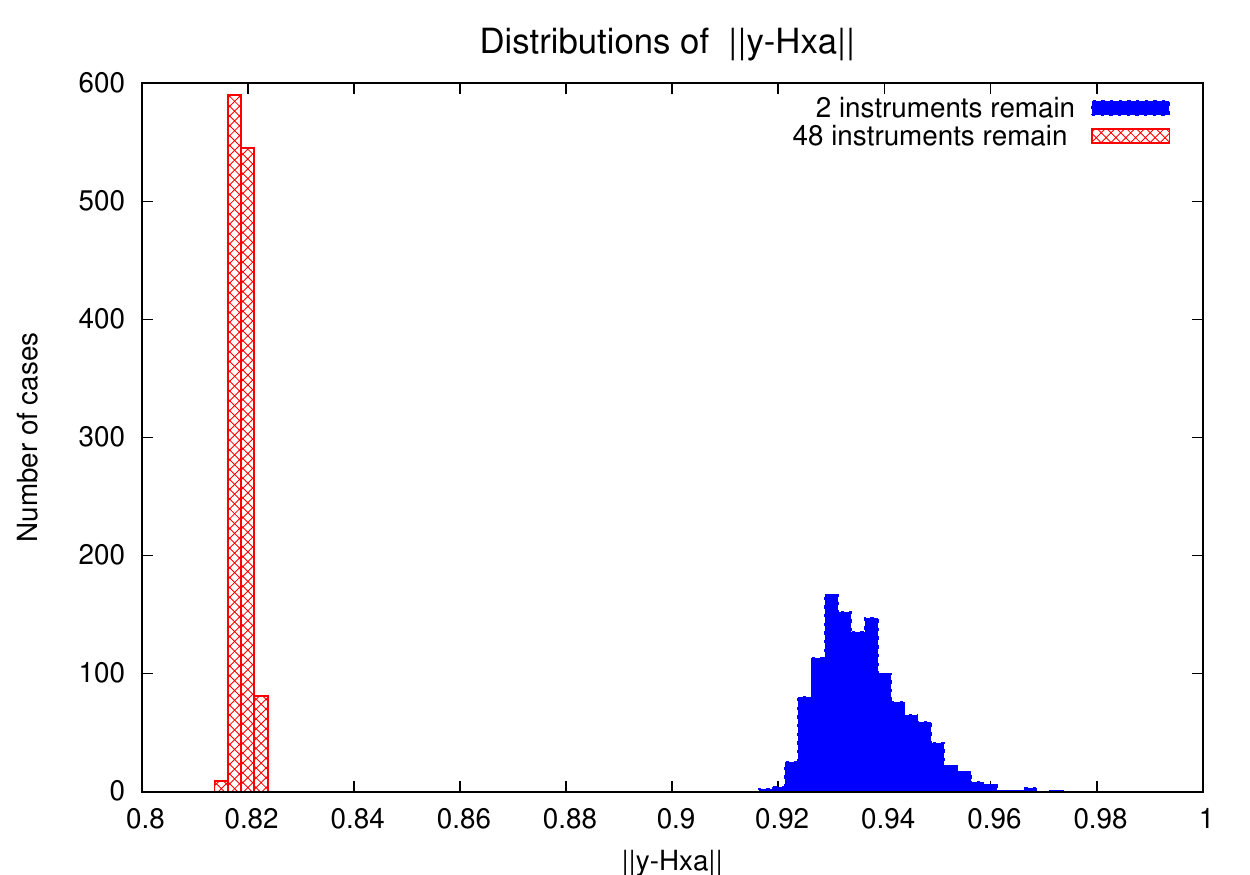} 

  \caption{Distribution of the norms of  $||\yo-\oH\xa||$ for all the cases
  where only two instruments are suppressed or two instruments remain in the
  instrument network on a PWR900 reactor. \label{fig:inf-fig1}}

\end{center}
\end{figure}

First, we explain why the third phase is marked by a stagnation of the mean
value of $||\yo-\oH\xa||$ over the set of removal scenarios taken into account.
To understand that effect, we  work on both the cases where two instruments are
removed ({\it i.e} 48 are remain), and the ones where two instruments are
remaining. On those two cases, at most $1225$ scenarios are possibles,
corresponding to the $C^{50}_{2}$ combinations. Using the hybrid Schur method,
we can calculate all those cases with a rather cheap computing time to obtain a
good statistics. We plotted the distribution of the value of $||\yo-\oH\xa||$ 
over all the scenarios in Figure \ref{fig:inf-fig1}. 

On the Figure \ref{fig:inf-fig1} we notice a big difference between the two
distributions plotted, that correspond to the cases where two instruments are
removed ({\it i.e.} $48$ remain) or two remain. 

The shifting of the mean value between those two extreme cases is logical as
available information is dramatically changing. However, the shape of the
distribution is also vastly changing. We move from a very sharp distribution,
when $48$ instruments remain, to a rather broad one when two instruments
remain. 

The first interesting point of this shape change is that all the instruments do
not have the same influence on the activity field reconstruction. To understand
better this effect let's assume that all the instruments are equivalent. In this
case, as the number of scenarios present in the two distributions of Figure
\ref{fig:inf-fig1} are the same, only a shifting of the mean value should have
be seen. However we have not only a translation of the distribution but also a
broadening. Thus, there is a non equivalence of instruments within the data
assimilation procedure, in terms of marginal information assigned to each
instrument, depending on its location in the core

The second point of interest is that the broadening of the distribution is
asymmetric. The distribution extends towards the higher values of norm. This
effect explains the stagnation of the $v$ quantity when few instruments
remain. The source of stagnation is the discrepancy between the most probable
value of the distribution and the mean value of the distribution which is
higher. The most probable value leads to a decrease without stagnation. However,
looking at the mean value (that has more physical meaning), we see that this one
stagnates due to the asymmetric broadening that compensates the overall decrease
in mean value.

Now the origin of the two slopes in decreasing of the information represented by
the Figure \ref{fig:pgi-fig1} will be investigated. The repartition of the
instruments in a standard PWR900 is very complex as shown in Figure
\ref{fig:figMFC}. This complexity of the repartition does not make the situation
easy to understand. Thus, we want to study the case on a simpler repartition of
the instrumentation to see if the effect of two phases decreasing persists.  

\section{Repartition effects in instrument removal}  

\begin{figure}[!ht]
\begin{center}
  \includegraphics[width=1.0\textwidth]{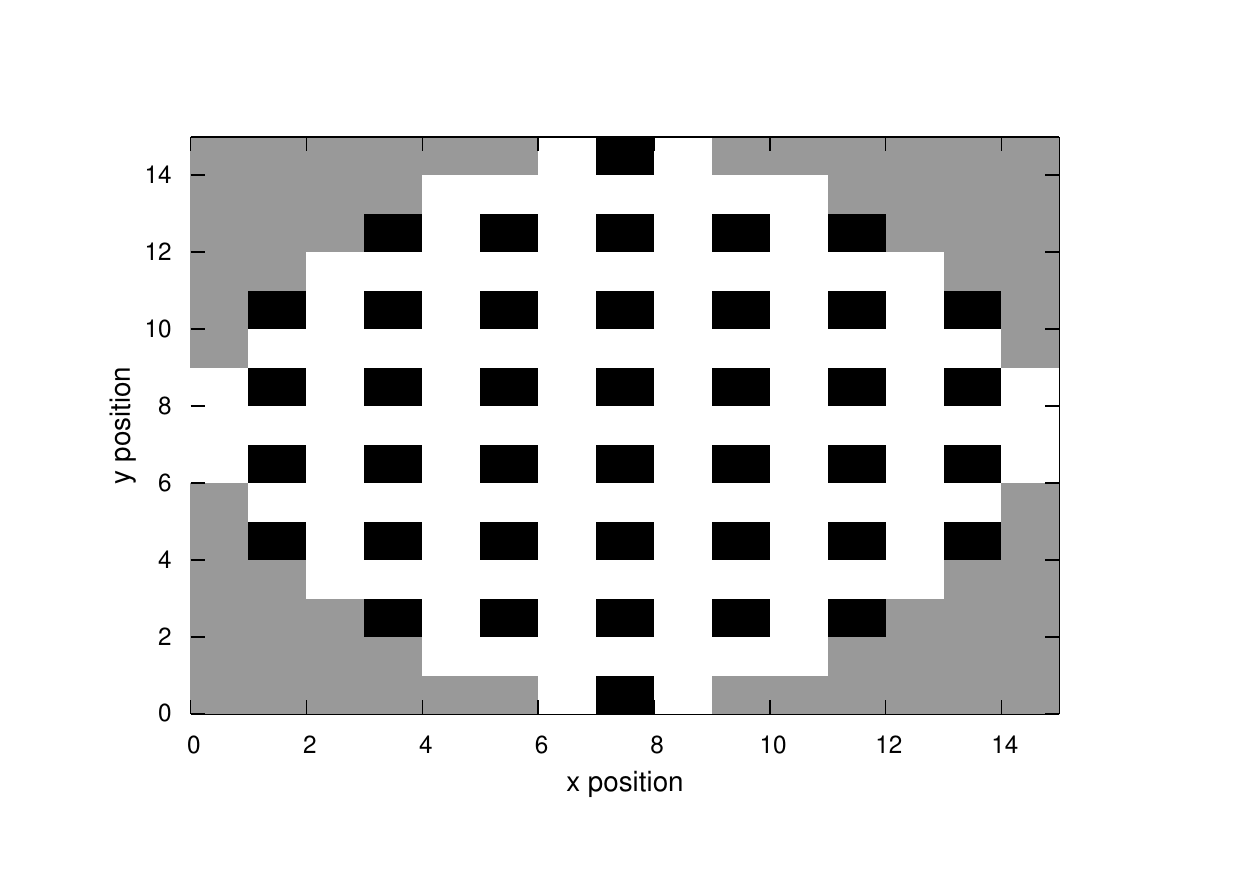}

  \caption{The MFC instruments within the nuclear core are localised in
  assemblies in black within the horizontal slice of the core. The assemblies
  without instrument are marked in white  and the reflector is in gray.
  \label{fig:pgi-figReg}}

\end{center}
\end{figure}

The position of the instruments are presented in Figure \ref{fig:pgi-figReg}.
The core geometry and assemblies configuration is the same as a PWR900 one,
however the instrument are located regularly on a Cartesian map.

\begin{figure}[!ht]
\begin{center} 
  \includegraphics[width=1.0\textwidth]{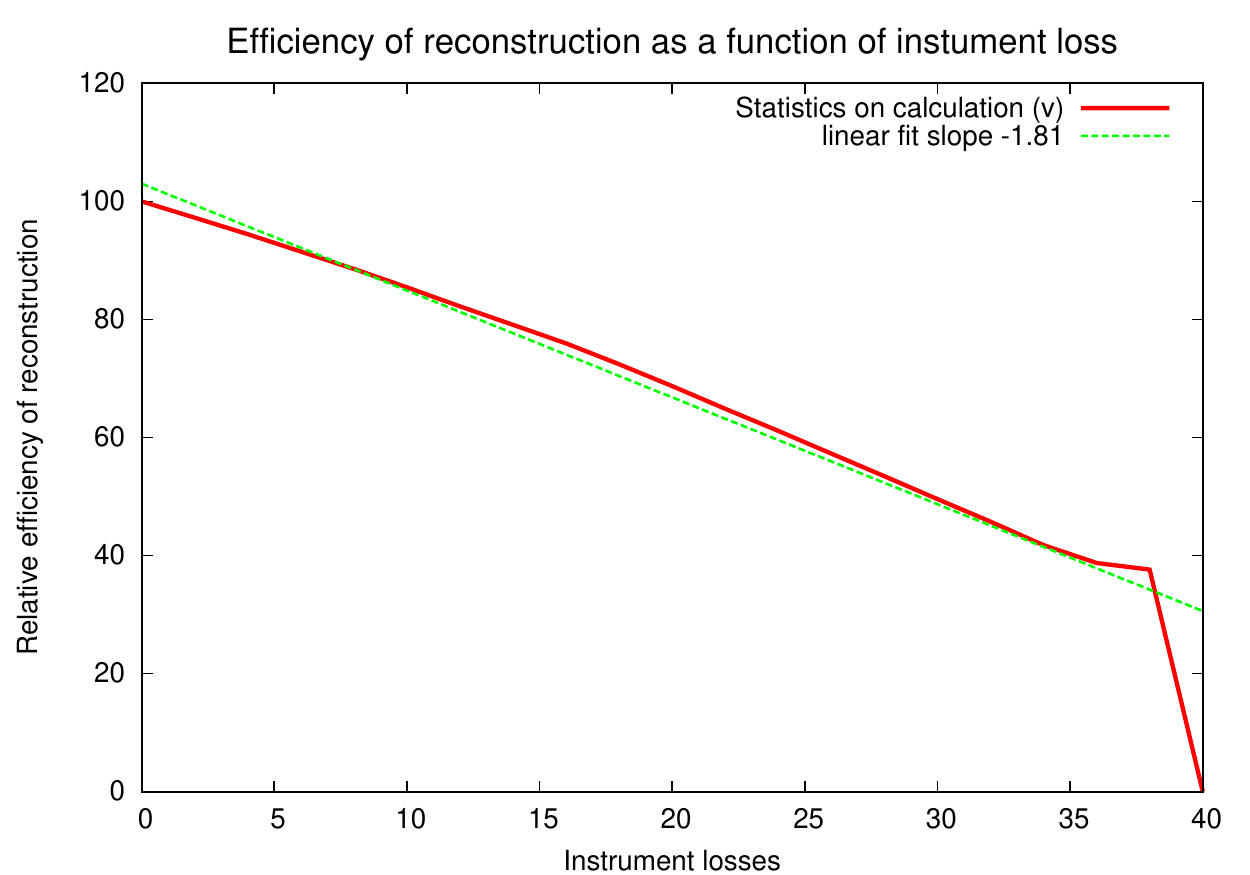}

  \caption{The plain curve represents the value $v$ given by Equation 
  \ref{yHxayHxb} as a function of the number of instruments removed. This
  results come from a regular instruments location in a PWR900. The dashed line
  represents the linear fit of the steady part of the curve.
  \label{fig:pgi-fig4}}

\end{center}
\end{figure}

Within this configuration, only $40$ MFC are used, which is a bit less than $50$
of the standard PWR configuration presented on Figure \ref{fig:figMFC}. With
this repartition, we do the same analysis as on the previous one. The evolution
on $v$ as a function of the number of removed instruments is plotted in Figure
\ref{fig:pgi-fig4}.

In the Figure \ref{fig:pgi-fig4}, the quality of the activity reconstruction 
decreases quasi-linearly as a function of the number of removed instruments.
This goes on until we reach the stagnation phase. It appears, comparing Figure
\ref{fig:pgi-fig4} to \ref{fig:pgi-fig1} that the variation of the decrease
slope is related to the geometrical repartition of the instruments. Such uniform
linear decreasing can be observed also in several other rather geometrically
regular repartition, as one with repartition on diagonal line. Some of the
repartitions have densities of instruments close to the one of the standard
PWR900, which do not change the overall behaviour. 

Looking at the slope factor of the linear fit, we notice that the one with
regular MFC repartition (Figure \ref{fig:pgi-fig4}) is in the range of the
slopes obtained  with standard MFC repartition (Figure \ref{fig:pgi-fig1}).

Thus the removal or fault resilience of the instrument set seems to depend on
the transition point between the two decreasing steps. In the lower range of
instrumental density, it is better to have a regular repartition, and in the
higher one, it is better to have a complex ad-hoc repartition. In real PWR900
nuclear core, because the set of intruments is fixed at a high density level,
these results indicate that it is more robust to have an ad-hoc repartition of
the instruments as for now.

\section{Conclusion}

We proposed and studied here an original method to test how the neutronic
activity field reconstruction by data assimilation is tolerant to information
removal. The core of the reconstruction method is based on data assimilation,
that is widely use in earth sciences, and allows a very good reconstruction of
the activity all over the nuclear core. An hybrid method for fast matrix
inversion partially based on Schur complement allows the execution on numerous
analysis from which statistical results are derived. For all these analyses,
synthetic data are used to try non-experimental intruments repartitions, but
similar results were established using real data. This application on real data
prove both the reliability and the quality of the calculation code and the data
assimilation methodology.

Using those advanced calculation methods, it was shown that the slopes of the
reconstruction quality  is mainly governed by repartition for the instruments.
Depending on the chosen repartition, the decrease consists in two or three
distinct phases. The ultimate stagnation phase in this decreasing is governed by
both statistical effect and heterogeneity of instruments influence.

The behaviour with two phases within the decreasing quality of the
reconstruction as a function of the number of instruments removed is understood
in term of repartition effect, but not quantified. However it can be seen as a
phase transition between two states of instrumental configuration. The
quantification of this transition is worth studying. 

\bibliography{bibliographie}

\begin{thebibliography}{10}

\bibitem{Massart07}
S\'ebastien Massart, Samuel Buis, Patrick Erhard, and Guillaume Gacon.
\newblock Use of \uppercase{3DVAR} and \uppercase{K}alman filter approaches for
  neutronic state and parameter estimation in nuclear reactors.
\newblock {\em Nuclear Science and Engineering}, 155(3):409--424, 2007.

\bibitem{Parrish92}
David~F. Parrish and John~C. Derber.
\newblock The national meteorological center's spectral statistical
  interpolation analysis system.
\newblock {\em Monthly Weather Review}, 120:1747--1763, 1992.

\bibitem{Todling94}
Ricardo Todling and Stephen~E. Cohn.
\newblock Suboptimal schemes for atmospheric data assimilation based on the
  kalman filter.
\newblock {\em Monthly Weather Review}, 122:2530--2557, 1994.

\bibitem{Ide97}
Kayo Ide, Philippe Courtier, Michael Ghil, and Andrew~C. Lorenc.
\newblock Unified notation for data assimilation: operational, sequential and
  variational.
\newblock {\em Journal of the Meteorological Society of Japan},
  75(1B):181--189, 1997.

\bibitem{era40}
S.~M. Uppala and {\it et al.}
\newblock The \uppercase{ERA}-40 re-analysis.
\newblock {\em Quaterly Journal of the Royal Meteorological Society}, 131(612,
  Part B):2961--3012, 2005.

\bibitem{Kalnay96}
E.~Kalnay and {\it et al.}
\newblock The \uppercase{NCEP/NCAR} 40-year reanalysis project.
\newblock {\em Bulletin of American Meteorological Society}, 77:437--471, 1996.

\bibitem{Huffman97}
G.~J. Huffman and {\it et al.}
\newblock The global precipitation climatology project (\uppercase{GPCP})
  combined precipitation dataset.
\newblock {\em Bulltin of American Meteorological Society}, 78:5--20, 1997.

\bibitem{Talagrand97}
Olivier Talagrand.
\newblock Assimilation of observations, an introduction.
\newblock {\em Journal of the Meteorological Society of Japan},
  75(1B):191--209, 1997.

\bibitem{Kalnay03}
Eugenia Kalnay.
\newblock {\em Atmospheric Modeling, Data Assimilation and Predictability}.
\newblock 2003.

\bibitem{Bouttier99}
François Bouttier and Philippe Courtier.
\newblock Data assimilation concepts and methods.
\newblock Meteorological training course lecture series, ECMWF, March 1999.

\bibitem{Rabier2000}
F.~Rabier, H.~{J\"arvinen}, E.~Kilnder, J.F. Mahfouf, and A.~Simmons.
\newblock The \uppercase{ECMWF} operational implementation of four-dimensional
  variational assimilation. part \uppercase{I}: Experimental results with
  simplified physics.
\newblock {\em Quarterly Journal of the Royal Meteorological Society},
  126:1143--1170, 2000.

\bibitem{Duderstatd76}
James~J. Duderstadt and Louis~J. Hamilton.
\newblock {\em Nuclear reactor analysis}.
\newblock John Wiley \& Sons, 1976.

\bibitem{Matheron70}
Georges Matheron.
\newblock {\em La th\'eorie des variables r\'egionalis\'ees et ses
  applications}.
\newblock Cahiers du Centre de Morphologie Math\'ematique de l'ENSMP,
  Fontainebleau, Fascicule 5, 1970.

\bibitem{Marcotte08}
Denis Marcotte.
\newblock G\'eologie et g\'eostatistique mini\`eres (cours), 2008.

\bibitem{SFP07}
Jean-Philippe Argaud, Bertrand Bouriquet, Patrick Erhard, Guillame Gacon, and
  S\'ebastien Massart.
\newblock Exploitation optimale des mesures neutroniques pour l'\'evaluation de
  l'\'etat des coeurs de centrales nucl\'eaires, \uppercase{SFP} conference,
  9-13 july 2007, 2007.

\bibitem{ecmi2008}
Jean-Philippe Argaud, Bertrand Bouriquet, Patrick Erhard, S\'ebastien Massart,
  and Sophie Ricci.
\newblock Data assimilation in nuclear power plant core.
\newblock In {\em \uppercase{ECMI} Conference, London, 30 June-4 July 2008}.
  IMA, 2008.

\bibitem{Zhang05}
Fuzhen Zhang.
\newblock {\em The \uppercase{S}chur complement and its applications}.
\newblock Springer, 2005.

\end{thebibliography}
\bibliographystyle{elsarticle-num}

\appendix
\section{Appendix: Schur complement method to optimise calculation\label{sec:schur}}  

Within the BLUE assimilation  method, the limiting factor in calculation time is
the matrix inversions. In Equation \ref{Kmat}, the costly part is the inversion
of the term:
\begin{equation} \label{Mmat}
  \oM = \oH \oB \oH^T +\oR.
\end{equation}
The inversion cost on huge matrix as $\oM$ (around $4000\times 4000$ in the
present case) was such that the time calculation of the above evaluation was
extremely time consuming. Then we had to optimise the computing cost. 

We noticed that the calculations are more time consuming when only few
instruments are removed. In this case the $\oM$ matrix is still huge. 

Thus, the idea is to use the information obtained in the inversion of the
full size matrix to shorten calculation, to calculate smaller size matrix in a
reasonable time. In this case, we want to calculate the new matrix as a
perturbation of the original one. Such a method exists and exploits
the Schur complement of the matrix. 

We assume we want to suppress some instruments to a given configuration. With
respect to the Equation \ref{Kmat}, we need to calculate a new matrix $\oK_n$.
The $n$ index is standing for referring the new matrix we want to calculate. For
that according to Equation \ref{Mmat} we have to determine a new matrix
$\oM_n$. 

This determination of $\oM_n$ is obtained from the knowledge of the invert of
the matrix $\oM_g$ calculated over all the instruments. The indices $g$ is used
to denote the reference matrix we start from $\oM_g$ according to Equation
\ref{Mmat}.

All the components of the new matrix $\oK_n$ can be obtained by suppressing the
lines and columns corresponding to removed instruments in $\oM_g$, inverting it
and then multiplying this matrix by the corresponding $\oH_n$ and  $\oB_n$. We
notice that, in our case, we get $\oB_n=\oB_g$ as we do not affect the model
space.

To make the demonstration easier, but without losing any generality, we can
assume that the suppressed instruments correspond to the lower square of 
$\oM_g$. If it is not the case, it is always possible to reorganise
matrix in such a way. 

Now we put the  $\oM_g$ under a convenient form, separating remaining
measures from removed ones. Assuming the starting matrix $\oM_g$ is  $m
\times m$ and that we plan so suppress $s$ measurements, we can write 
$\oM_g$ in the following way:
\begin{equation}   \label{pgi:dec}
\oM_g =
\left(
\begin{array}{cc} 
\oP_g & \oQ_g  \\ 
\oR_g & \oS_g  
\end{array} 
\right),
\end{equation}
where:
\begin{itemize} 

\item $\oP_g$ contains the remaining measurements, and is a $p \times p$ matrix,

\item $\oS_g$ contains the suppressed measurement, and is a $s \times s$ matrix,

\item$\oQ_g$ et $\oR_g$ represent the dependence between remaining measured and
suppressed ones. In the particular case we are dealing with, we have
$\oQ_g^T=\oR_g$. However, no further use of this property is done.

With such a decomposition, we got the equality $m=p+s$.

\end{itemize}
The  $\oP_g$  matrix corresponds to the remaining instruments, thus we
have the equality:
\begin{equation} \label{pgi:Pmat}
\oP_g  = \oM_n,
\end{equation} 
The decomposition given in Equation \ref{pgi:dec} is the one required to build
the Schur complement of this matrix \cite{Zhang05}. Under the condition that
$\oP_g$ can be inverted, the Schur complement is the following quantity:
\begin{equation} \label{pgi:shur}
\oS_g - \oR_g \oP_g^{-1} \oQ_g,
\end{equation} 
and is noted $(\oM_g/\oP_g)$. This notation reads as Schur complement of $\oM_g$
by $\oP_g$.

Thus we look for a cheap way to calculate $\oP_g^{-1}$ knowing $\oM_g^{-1}$. For
that, we use the  Banachiewiz formula \cite{Zhang05} that gives invert of
$\oM_g$  as a function of $\oP_g$, $\oQ_g$, $\oR_g$, $\oS_g$ and $(\oM_g/\oP_g)$
matrices:
\begin{eqnarray}  \label{pgi:banach}
\oM_g^{-1} & = &
\left (
\begin{array}{cc} 
\oP_g & \oQ_g  \\ 
\oR_g & \oS_g  
\end{array} 
\right )^{-1}  
\\
 & = & \left (
\begin{array}{cc} 
\oP_g^{-1} + \oP_g^{-1} \oQ_g (\oM_g/\oP_g)^{-1} \oR_g \oP_g^{-1} & -\oP_g^{-1} \oQ_g (\oM_g/\oP_g)^{-1} \\ 
-(\oM_g/\oP_g)^{-1} \oR_g \oP_g^{-1} & (\oM_g/\oP_g)^{-1}
\end{array} 
\right ) \nonumber
\end{eqnarray}  
We define the 4 sub-matrices $\tP_g$, $\tQ_g$, $\tR_g$ and $\tS_g$ by:
\begin{eqnarray} 
\tP_g & = & \oP_g^{-1} + \oP_g^{-1} \oQ_g (\oM_g/\oP_g)^{-1} \oR_g\oP_g^{-1},\\
\tQ_g & = & -\oP_g^{-1} \oQ_g (\oM_g/\oP_g)^{-1},\\
\tR_g & = & -(\oM_g/\oP_g)^{-1} \oR_g \oP_g^{-1},\\
\tS_g & = & (\oM_g/\oP_g)^{-1}.
\end{eqnarray} 
Rearranging those terms we get:
\begin{equation} 
\oP_g^{-1} = \tP_g - \tQ_g \tS_g^{-1} \tR_g.
\end{equation} 
As by hypothesis we know the inverse $\oM_g^{-1}$ of the global matrix, we are
able to extract $\tP_g$, $\tQ_g$, $\tR_g$ and $\tS_g$ from the whole inverted
matrix. Thus, the main cost to obtain the inverse of $\oP_g$  of size $p \times
p$ becomes the one of inverting  $\tS_g$  which size is $q \times q$. In first
approximation,  if the number of measurements to suppress is smaller than the
number of remaining one, this methods gives a notable gain. As soon as the
matrix $\oP_g^{-1}=\oM_n^{-1}$, final calculation of $\oK_n$ is straightforward.

\begin{figure}[!ht]
\begin{center}
 \includegraphics[width=1.0\textwidth]{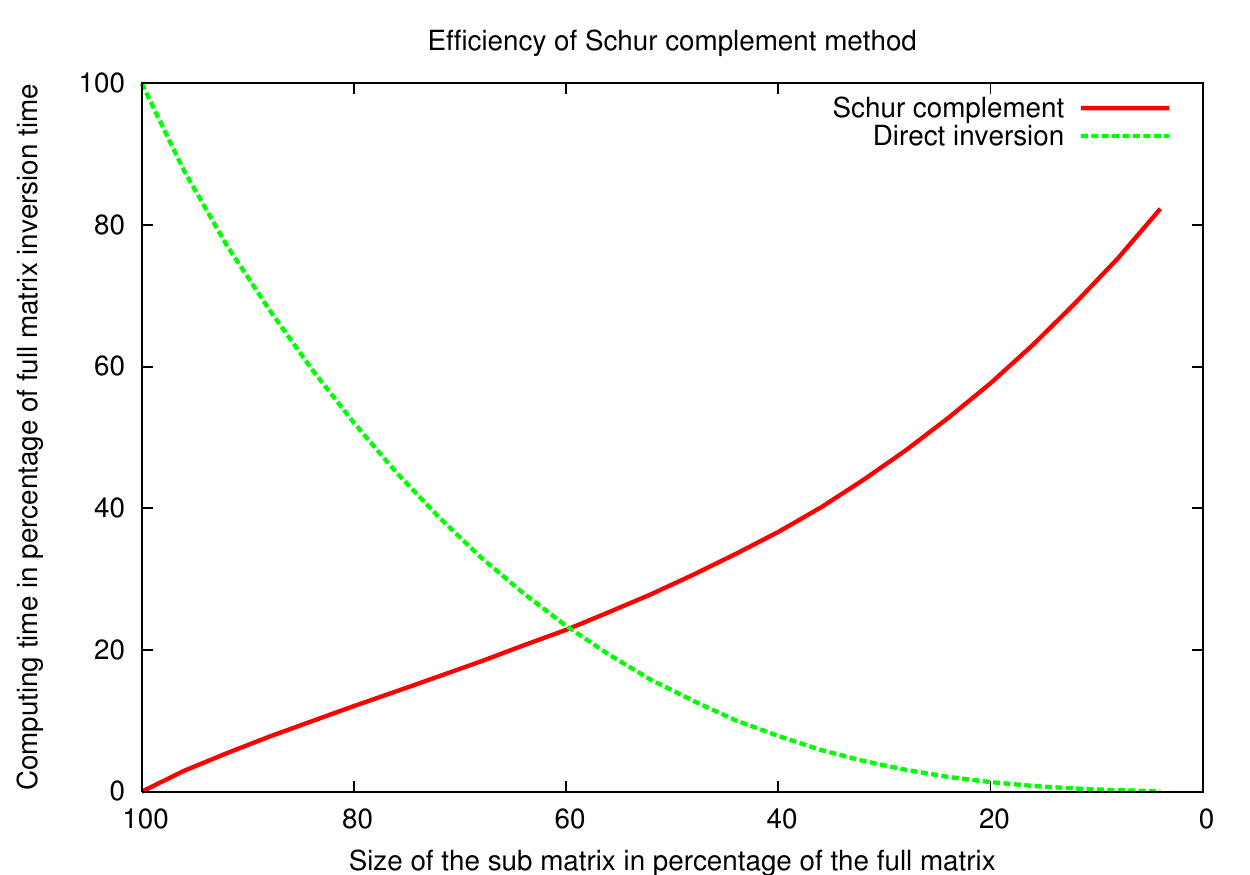}

 \caption{Calculation time as a function of the size of the sub-matrix.
 Calculation time is given as a percentage of the calculation time of the full
 size matrix. The size of the sub-matrix is given a a percentage of the size of
 the full matrix. The curves in full line (red) and in dashed line (green)
 represent the inversion by Schur complement and direct inversion respectively.
 \label{fig:pgi-fig0}}

\end{center}
\end{figure}

To highlight the advantages of this method with respect to the standard
inversion of sub-matrix, some tests are shown on a $4000 \times 4000$ full
semi-definite positive regular matrix. The curves showing the effective
computing time in percentage of the computing time of the full matrix are
presented in Figure \ref{fig:pgi-fig0}.

Figure \ref{fig:pgi-fig0} shows that, when sub-matrix has roughly the size of
the initial matrix, the inversion by Schur complement is far more efficient than
the direct inversion. Above around  $60\%$ of the size of the initial matrix,
the direct inversion becomes more efficient. The crossing point is at $60\%$ 
instead of $50\%$, as expected in first approximation. This difference comes
from the few additional multiplications that need to be done in the Schur
complement calculation, as we can see in Equation \ref{pgi:shur}. Globally, we
see that the most efficient method is to use an hybrid calculation that choose
the best way to make the calculation as a function of the number of measurements
removed. To quantify the improvement of such an hybrid choice, we integrate the
curves of direct inversion and compare it to the integral of hybrid option
(minimum of both cases) within the instrument loss range. The ratio in
percentage of both integrals shows that benefit of this hybrid method represents
an overall gain of $64\%$ with respect to standard method. 

\end{document}